\documentclass{evn2002}
\usepackage{natbib}
\usepackage{graphicx}
\begin{document}
   \title{The Relative Spatial Distribution of SiO Masers in AGB Stars
   at 43 and 86 GHz }

   \author{J.-F. Desmurs\inst{1}, R. Soria-Ruiz\inst{1}, F.
          Colomer\inst{1}, K.B. Marvel\inst{2}, V.
          Bujarrabal\inst{1}, J. Alcolea\inst{1}, P.J.
          Diamond\inst{3}, D. Boboltz\inst{4} \and A. Kemball\inst{5}
          }

   \institute{O.A.N, Apartado 1143, Alcal\'a de Henares, Spain
         \and
             American Astronomical Society, USA
         \and
             Jodrell Bank Observatory, U.K.
         \and 
             USNO, USA
         \and
             NRAO, Socorro, New Mexico, USA
             }

   \abstract{ We present the first VLBI images of SiO masers in the
circumstellar envelope of an S-type star obtained with the newly
developed capabilities of the Very Long Baseline Array (VLBA) at 86
GHz. We combine these data with those obtained quasy simultaneously at
43 GHz.  These observations provide information on the structure and
dynamics of the innermost circumstellar shells, where the return of
large quantities of stellar material to the interstellar medium starts.
Despite of the fact that these are prelimininary results, we report
that the maser emission of the $v$=1 J=1--0 and J=2--1 present a very
different emission distribution, being not coincident in most cases.  }
\authorrunning{J.-F Desmurs et al.}  
\maketitle
%
%________________________________________________________________

\section{Introduction}

SiO masers emission at 7 mm wavelength ($v$=1 and $v$=2, J=1--0
transition near 43 GHz) has been observed in AGB stars with very hight
resolution by means of VLBI techniques, yielding important results in
relation with their not yet well understood pumping mechanism. The 7 mm
maser emission regions are found to be distributed in a number of spots
forming a ring-like structure at about 2-3 stellar radii
\citep{col92,dia94,gre95,des00} which are assumed to be centered on the
stellar position.  This ring-like flux distribution arise naturally in
the framework of the radiative pumping mechanism of SiO masers (see
Bujarrabal et al. 1994), but may also be explained by a collisional
modelisation.  Recent simultaneous observations of the $v$=1 and $v$=2,
J=1--0 (see Desmurs et al. 2000) show that the $v$=1 and $v$=2 maser
spots are often close, but appear systematically shifted by a few mas
and are only rarely coincident; a result that would argue in favor of
radiative pumping models.

In an attempt to pursue the comparison between the predictions of the
different models and observational data, we have measured the relative
spatial distribution of SiO maser emission between the J=1--0 and
J=2--1 transitions. In that case, models using either radiative or
collisional pumping predict ``maser chains'' across the excitated
vibrational states, in such a way that the inversion of the different-J
transition in the same $v$ state are mutually reinforced.  Models
predict that the rotational masers in the same vibrational state should
appear under the same physical conditions and, therefore, that the
43\,GHz and 86\,GHz masers emission originate from the same
condensations in the circumstellar envelope (CSE)
\citep[see][]{buj94,hum02}.

\section{Observations and data analysis}

\begin{figure}[t]
 \centering 
 \includegraphics[scale=0.45]{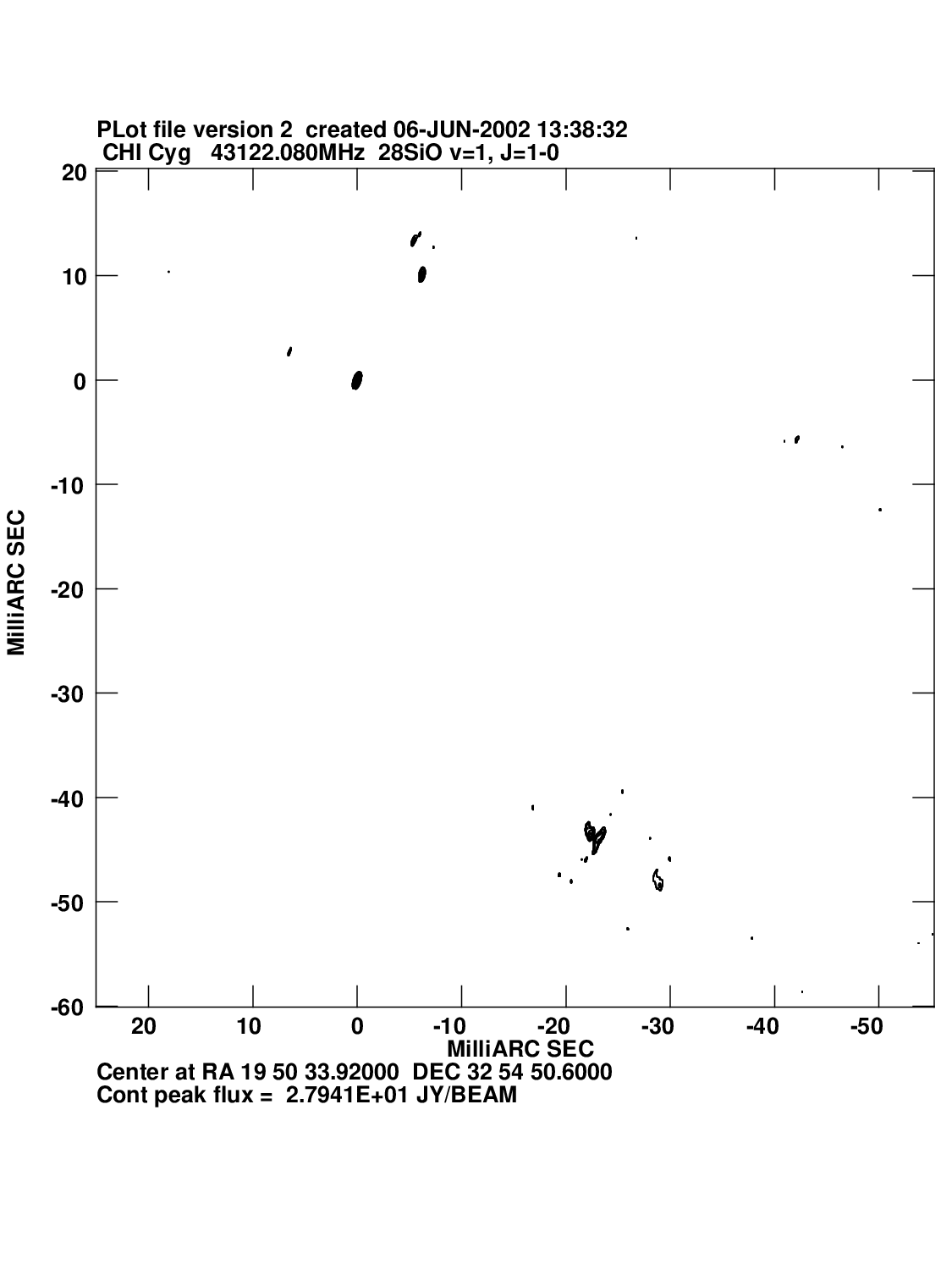}
 \vspace{-1cm}
 \caption{Integrated intensity maps of the $v$=1, J=1--0 line (rest
frequency 43122.080 MHz) of SiO masers towards $\chi$ Cyg. Contours are
multiples by 10\% of the peak flux ($\sim$28 Jy). The resolution beam is
0.8x0.21 mas (PA=-12.4$^{\circ}$). }
\label{chi_cyg_7mm}
\end{figure}

We have performed quasi-simultaneous observations (separated only by
few hours) of the J=1--0 $v$=1 and $v$=2 (at 7mm), J=2--1 $v$=1 and
$v$=2 lines of $^{28}$SiO, and the J=1--0 $v$=0 line of $^{29}$SiO (at
3mm) with NRAO Very Long Baseline Array, on May 9th, 2001 of three AGB
stars: TX Cam, R Cas and $\chi$ Cyg.  This was possible with the new
capabilities of the VLBA at 3 mm.  The system was setup to record 4 MHz
at 7\,mm in dual polarization and 8 MHz at 3\,mm in single
polarization. The correlation was produced at the VLBA correlator in
Socorro (NM, USA) providing 256 and 512 spectral channels respectively
for a final spectral resolution of $\sim$0.1 and
$\sim$0.05\,km~s$^{-1}$.  The calibration was performed using the
standard scheme in the Astronomical Image Processing System (AIPS) for
spectral line experiments; the amplitude calibration was performed
applying the template spectra method. In this paper, we report only on
the first results we obtained for the source $\chi$~Cyg in the $v$=1,
J=1--0 (at 43122.080~MHz see Fig~\ref{chi_cyg_7mm}) and $v$=1, J=2--1
(at 86243.442~MHz see Fig~\ref{chi_cyg_3mm}) lines. Maps of the two
transitions were independently produced by solving the residual
fringe-rates on the line source. This correction is determined by
selecting a channel containing a simple feature with a simple structure
and a high signal to noise ratio, that is used as phase reference for
all other channels.  This fixed the position of that maser spot as the
map origin, in both cases.

\section{Results and discussion}

\begin{figure}[t]
 \centering
 \includegraphics[scale=0.45]{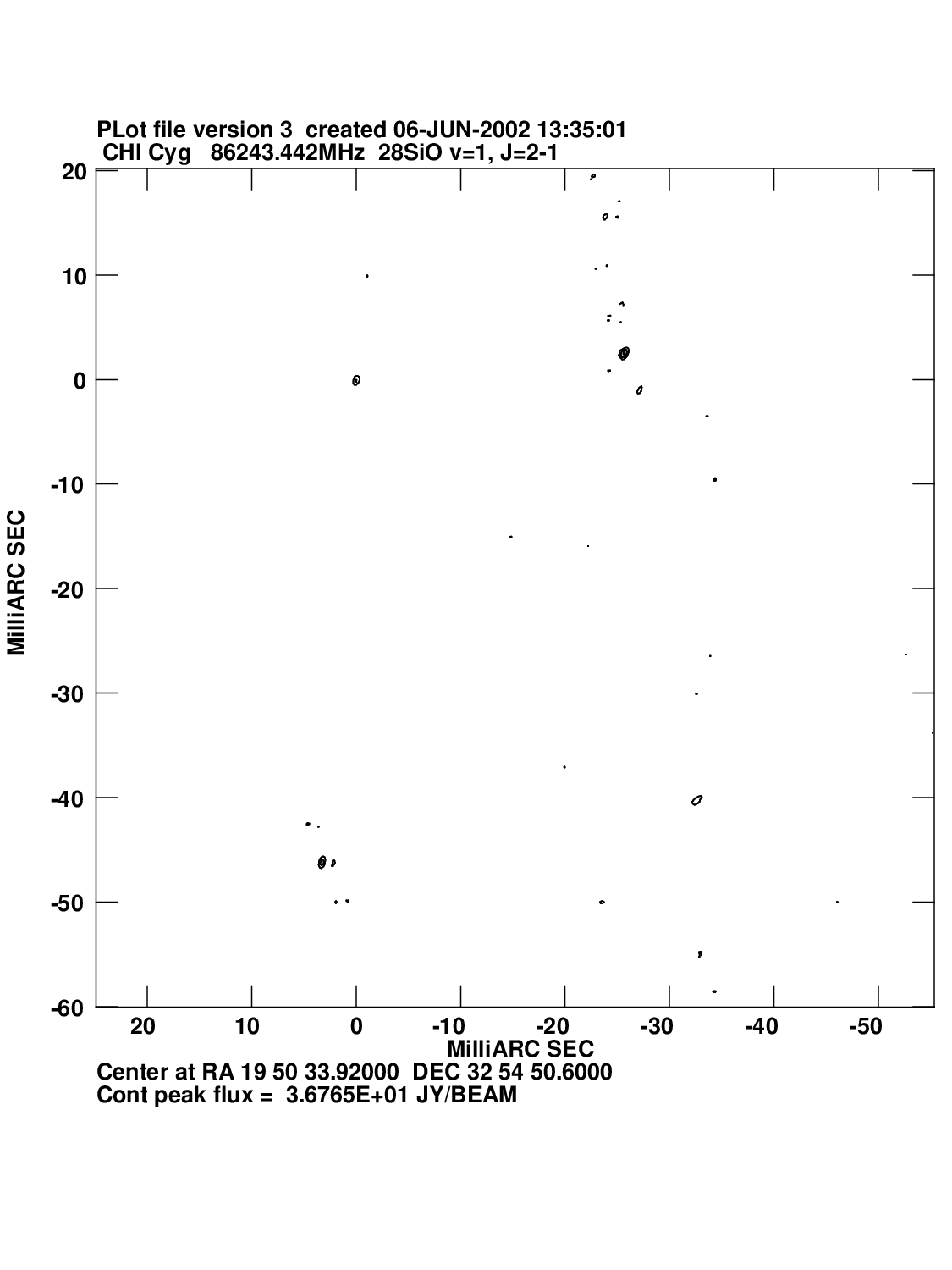} 
 \vspace{-1cm}
 \caption{Integrated intensity maps of the $v$=1, J=2--1 line (rest
frequency 86243.442 MHz) of SiO masers towards $\chi$ Cyg. Contours are
multiples by 10\% of the peak flux ($\sim$36 Jy). The resolution beam is
0.75x0.052 mas (PA=-15.8$^{\circ}$). }
\label{chi_cyg_3mm}
\end{figure}

We produced the first VLBA maps with milliarcsecond resolution of the
SiO maser emission from an S-type star, $\chi$~Cyg.  We obtained maps
of the transitions $v$=1, J=1--0 at 7\,mm (see Fig \ref{chi_cyg_7mm})
and of the $v$=1, J=2--1 at 3\,mm (see Fig \ref{chi_cyg_3mm}).  The map
size is approximately $\sim$80 by 80 mas. The gaussian restoring beam
has a FWHM of 0.8x0.2\, mas (with a position angle PA=-12.4$^{\circ}$)
and 0.75x0.05\, mas (with a position angle PA=-15.8$^{\circ}$),
respectively, for the data at 7 and 3\,mm. Both figures are using the
same scale to ease a direct comparison, but do not share spatial origin
as explained before.

Our preliminary results shows that the SiO maser emission distribution
in both transition occur more or less at a similar radii from the
center.  In particular, we observed that masers for $v$=1, J=2--1
(3\,mm) arise from a ring-like structure, as it has been reported in
several other AGB stars at 43 GHz \citep{dia94, gre95, des00, phi01}.
The radii of this ring structure is of the order of $\sim$ 28\,mas
which, adopting the Hipparcos distance (ESA 1997) of 106$\pm$15 pc, is
equivalent to 4.5\,10$^{13}$\,cm.
% equivalent a 2.97 AU

One of the most surprising result of these observations is that the
emission distribution between the two lines is completly
different. Whatever the choosed alignment, it is impossible to make
coincident more than one maser spot at the same time. Even the regions
emitting in one transition or the other are very different at large
scale (see Figs~\ref{chi_cyg_7mm} and~\ref{chi_cyg_3mm}). This is in
complete contradiction with all theoretical predictions.

Our conclusions are only preliminary. But in the case that they would
be confirmed by future observations and in other sources, this result
would cast serious doubts on the present models of the SiO maser
excitation.


\begin{thebibliography}{}

\bibitem[Bujarrabal et al.(1994)]{buj94} Bujarrabal, V., 1994, 
A\&A 285, 953

\bibitem[Colomer et al.(1992)]{col92} Colomer, F., Graham, D. A.,
Krichbaum, T. P., Ronnang, B. O., de Vicente, P., Witzel, A., Barcia,
A., Baudry, A., Booth, R. S., Gomez-Gonzalez, J., Alcolea, J., Daigne,
G, 1992, A\&A 254, L17

\bibitem[Diamond et al.(1994)]{dia94}
Diamond, P., Kemball, A. J., Junor, W., Zensus, A., Benson, J., Dhawan,
V., 1994, ApJ Letter 430, L61

\bibitem[Desmurs et al.(2000)]{des00}
Desmurs, J.-F., Bujarrabal, V., Colomer, F., \& Alcolea, J., 2000, A\&A
360, 189

\bibitem[Greenhill et al.(1995)]{gre95}
Greenhill,L.J, Colomer, F., Moran, J. M., Backer, D. C., Danchi, W. C.,
Bester, M., 1995, ApJ 449, 365

\bibitem[Humphreys et al.(2002)]{hum02} 
Humphreys, E.M.L., Gray, M.D., Yates, J.A., Field, D., Bowen, G.H. \&
Diamond, P.J., 2002, A\&A 386, 256

\bibitem[Phillips et al.(2001)]{phi01}
Phillips, R.B., Sivakoff, G.R., Lonsdale, C.J., Doeleman, S.S.,
2001, AJ 122, 2674.

\end{thebibliography}
\end{document}